\documentclass[aps,prl,preprint,groupedaddress,showpacs]{revtex4-1}

\usepackage{graphicx}
\usepackage{amsfonts}

\newcommand{\im}{{\rm Im}}

\newcommand{\dd}{{\rm d}}

\begin{document}

\title{Scaling of the dynamics of homogeneous states of one-dimensional long-range interacting systems}
\author{T.~M.~Rocha Filho}
\affiliation{Instituto de F\'\i{}sica and International Center for Condensed Matter Physics\\ Universidade de
Bras\'\i{}lia, CP: 04455, 70919-970 - Bras\'\i{}lia, Brazil}
\author{A.~E.~Santana}
\affiliation{Instituto de F\'\i{}sica and International Center for Condensed Matter Physics\\ Universidade de
Bras\'\i{}lia, CP: 04455, 70919-970 - Bras\'\i{}lia, Brazil}
\author{M.~A.~Amato}
\affiliation{Instituto de F\'\i{}sica and International Center for Condensed Matter Physics\\ Universidade de
Bras\'\i{}lia, CP: 04455, 70919-970 - Bras\'\i{}lia, Brazil}
\author{A.~Figueiredo}
\affiliation{Instituto de F\'\i{}sica and International Center for Condensed Matter Physics\\ Universidade de
Bras\'\i{}lia, CP: 04455, 70919-970 - Bras\'\i{}lia, Brazil}
\begin{abstract}
Quasi-Stationary States of long-range interacting systems have been studied at length over the last fifteen years.
It is known that the collisional terms of the Balescu-Lenard and Landau
equations vanish for one-dimensional systems in homogeneous states, thus requiring a new kinetic equation with a proper dependence
on the number of particles. Here we show that the scalings discussed in the literature are mainly due either to small size effects or the use of
unsuitable variables to describe the dynamics. The scaling obtained from both simulations and theoretical considerations
is proportional to the square of the number of particles
and a general form for the kinetic equation valid for the homogeneous regime is obtained. Numerical evidence is given for
the Hamiltonian Mean Field and Ring models and a kinetic equation valid for the homogeneous state is obtained for the former system.
\end{abstract}

\maketitle

Classical systems with long range interaction can present unusual properties such as as
non-ergodicity, anomalous diffusion, aging, non-Gaussian Quasi-Stationary States, negative microcanonical heat capacity and ensemble inequivalence, captured from the well known result of the positivity of the heat capacity in the canonical ensemble ~\cite{physrep,proc1,proc2,proc3,dauxois,metodo,noneq,nos,eplnos,10b}.
A pair interaction potential is said to be long ranged if it decays at long distances
as $r^{-\alpha}$ with $\alpha \leq d$ where $d$ is the spatial dimension~\cite{dauxois}. The dynamics of such systems has essentially three stages:
(i) a violent relaxation ~\cite{lyndenbell,lb1} even though a satisfactory theory is still lacking~\cite{arad1,arad2,eplnos,debuyl}),
towards a Quasi Stationary State (QSS) in a short time roughly independent on the number of particles $N$;
(ii) a QSS with a very long relaxation time to thermodynamic equilibrium that diverges with $N$,
and finally, (iii) the system reaches the thermodynamic equilibrium. In the $N\rightarrow\infty$ limit the stage (iii) is never attained. After the first stage the system may also oscillate around a QSS with an amplitude decreasing with time
due to a non-linear Landau damping~\cite{mouhot}. The slowly varying state remains
Vlasov stable and in some cases may loose its stability
and rapidly evolve into another QSS, thereby resuming the slow evolution towards equilibrium~\cite{artigo1}.
This slow dynamics of the QSS have been extensively studied
in the literature (see~\cite{physrep,proc1,proc2,proc3,20h,20b,20c,20g,20d,20e,20f,20i} and references therein)
for different systems such as the Hamiltonian Mean Field (HMF)
model~\cite{nv1,nv2,nv3,jain,buyl,hmforig}:
\begin{equation}
H=\sum_{i=1}^N \frac{p_i^2}{2}+\frac{1}{N}\sum_{i<j=1}^N \left[1-\cos(\theta_i-\theta_j)\right],
\label{hmfham}
\end{equation}
with $\theta_i$ the position angle of particle $i$ on a circle and $p_i$ its conjugate momentum.
Here we also investigate the dynamics of the homogeneous QSS for the Ring model with Hamiltonian~\cite{sota}:
\begin{equation}
H=\sum_{i=1}^N \frac{p_i^2}{2}+\frac{1}{N}\sum_{i<j=1}^N \frac{1}{\sqrt{2}\sqrt{1-\cos(\theta_i-\theta_j)+\epsilon}},
\label{ringham}
\end{equation}
where $\theta_i$ and $p_i$ have the same meanings as for the HMF model.
Other systems with long-range interactions of interest, but not considered here, are one, two and
three-dimensional self-gravitating systems discussed in ~\cite{valageas,nv4,nv4b,nv5,nv6},~\cite{teles,marcos} ~\cite{chandrasekhar,heyvaerts,diemand,gabrielli}, respectively. The study of such systems along the same lines  will be addressed in a future publication.

The first stage of the dynamics is described by the Vlasov equation (VE) which is satisfied by the one particle distribution function
in the $N\rightarrow\infty$ limit~\cite{braun,artigo1}. For finite $N$ this equation is valid only for short times encompassing
the outburst of the violent relaxation. After this initial stage, collisional effects (graininess) accumulate and the VE must be corrected by considering
higher order terms in an expansion in powers of $1/N$ as it will be discussed below, leading to kinetic equations such as the Landau or Balescu-Lenard equations~\cite{balescu1,lenard,liboff}.

A sensible revision of the known kinetic equations for long-range interacting systems and their deductions, with all proper references,
is presented by Chavanis in Refs.~\cite{chavanis,chavanisa,chavanisb,chavanisc}.
These equations usually can be obtained from the BBGKY hierarchy~\cite{liboff} by taking into account contributions
from the two-body correlation functions, which are of order $1/N$~\cite{balescu1,chavanis} that result in a time scale of collisional relaxation proportional
to $N$~\cite{chavanis}. For three-dimensional gravity the dynamics scales as $N/\log N$, known as the Chandrasekhar scaling~\cite{chandrasekhar2}.
The Balescu-Lenard equation for a one-dimensional homogeneous system is written as~\cite{balescu1}:
\begin{equation}
\frac{\partial}{\partial t}f_1(p_1;t)=2\pi^2n\frac{\partial}{\partial p_1}\int {\rm d}p_2\int {\rm d}k
\frac{k^2\tilde V(k)^2}{\left|\varepsilon(k,kp_1) \right|^2}\delta(k(p_1-p_2))
\left(\frac{\partial}{\partial p_1}-\frac{\partial}{\partial p_2}\right)
f_1(p_1;t)f_1(p_2;t),
\label{ballen}
\end{equation}
where $f_1(p_1;t)$ is the one-particle reduced distribution function, $n$ the particle density, $p_i$ the momentum of particle $i$,
$\tilde V(k)$ the Fourier transform of the pair interaction potential and $\varepsilon(k,kp_1)$ is the dielectric function.
Collective effects are ruled out if one takes $\varepsilon(k,kp_1)=1$ and that results in the Landau equation.
The right-hand side of Eq.~(\ref{ballen}) vanishes identically due to the Dirac delta function~\cite{eldridge,kadomtsev}.
Therefore higher order terms must be kept
when truncating the hierarchy, leading to a different scaling of the time evolution of a homogeneous state.
More recently Sano proposed a
derivation of a kinetic equation for one-dimensional homogeneous systems by summing contributions of all orders in the hierarchy~\cite{sano}.
Unfortunately his approach is limited to dilute gases and is not relevant for the problems addressed in this paper.
It would be natural to expect that in the present case the predominant collisional corrections to the
kinetic equation come from higher order terms proportional to
$1/N^2$. This implies a relaxation scaling proportional to $N^2$,
as expected from a more straightforward application of kinetic theory such as that discussed by Chavanis
in Refs.~\cite{chavanis,chavanisa,chavanisb,chavanisc}, even though he did not obtained a closed form kinetic equation at this order.

The $N^2$ scaling of the dynamics has previously been observed for one-dimensional neutral plasmas~\cite{dawson,rouet}.
However, different scalings proportional to $N^{1.7}$ and
$\exp(N)$ were reported in Refs.~\cite{Zan},\cite{ nv1} and~\cite{nv2}, respectively, for different types of initial conditions. The present authors obtained for the HMF model and different type of initial condition
a scaling proportional to $N^2$ Ref.~\cite{artigo1}, which strongly suggests that the dominant contribution to the collisional term in the kinetic equation is
given by the next term in the $1/N$ expansion.
A possible explanation for these discrepancies may originate from the fact that the number of particles in the simulations described in~\cite{artigo1}
are much greater than in~\cite{nv1} and~\cite{nv2}, thereby a possible finite size effect should be carefully examined.

We argue that to properly probe how the dynamics depends on the number of particles
a better choice of dynamical variables is to use higher moments of the velocity distribution,
but keeping in mind that the second moment of $p$ is constant in the QSS due to energy conservation (homogeneity fixes the value of the magnetization and therefore
of the potential energy up to small fluctuations). On the other hand, both this moment and the total magnetization
can be used to determine the life-time of the homogeneous QSS given by the value of time at which their
previously constant values start to change rapidly.

We stress the point that the magnetization is not a useful variable to follow the system dynamics (and therefore its scaling with $N$),
as it depends only on the spatial distribution which is fixed in a homogeneous QSS.

In this paper one shows that the direct observation of the higher order moments of the velocity distribution
leads to a different estimation of the scaling of the dynamics. 
We extend the calculations for the Ring model and show that the $N^2$ scaling is observed for homogeneous states of the systems described by this model. In this case, up to the authors knowledge, the scaling
of the dynamics of the QSS for this model has not been studied due to the difficulty to pinpoint homogeneous QSS
with a finite lifetime and the computation cost necessary for numerical simulations.

To corroborate our arguments (i) we have performed numerical simulations for the HMF model using the same type of initial
conditions as considered in Refs.~\cite{nv1,nv2} and varying the number of particles up to larger values than those encompassed in previous studies and, (ii)  derive the kinetic equation to show that the relaxation scaling of the dynamics is proportional to  $N^2$ .

We note that the $\exp(N)$ scaling as claimed in Ref.~\cite{nv2} was obtained by conjecturing on the extrapolation
of the dynamics of the phase of the magnetization vector, though it lacks more formal justification.

As explained above $M_2=\langle p^2\rangle$ is constant (up to fluctuations) in a homogeneous QSS, so we look at the fourth moment
$M_4=\langle p^4\rangle$ of the momentum distribution starting from a homogeneous
waterbag initial condition defined by $f(p;0)=1/2p_0\:\:{\rm if}\:\:|p|<p_0$, and $0$ otherwise ($p_0$ constant).
The moment $M_4$ varies slowly with time as a consequence of collisional corrections to the VE, the dependence
of the dynamics on $N$ being the inverse of that for the collisional term in the kinetic equation.
All simulations were performed using a parallel implementation of a fourth-order symplectic
integrator as described in Ref.~\cite{eu2}.

\begin{figure}[ptb]
\begin{center}
\scalebox{0.3}{{\includegraphics{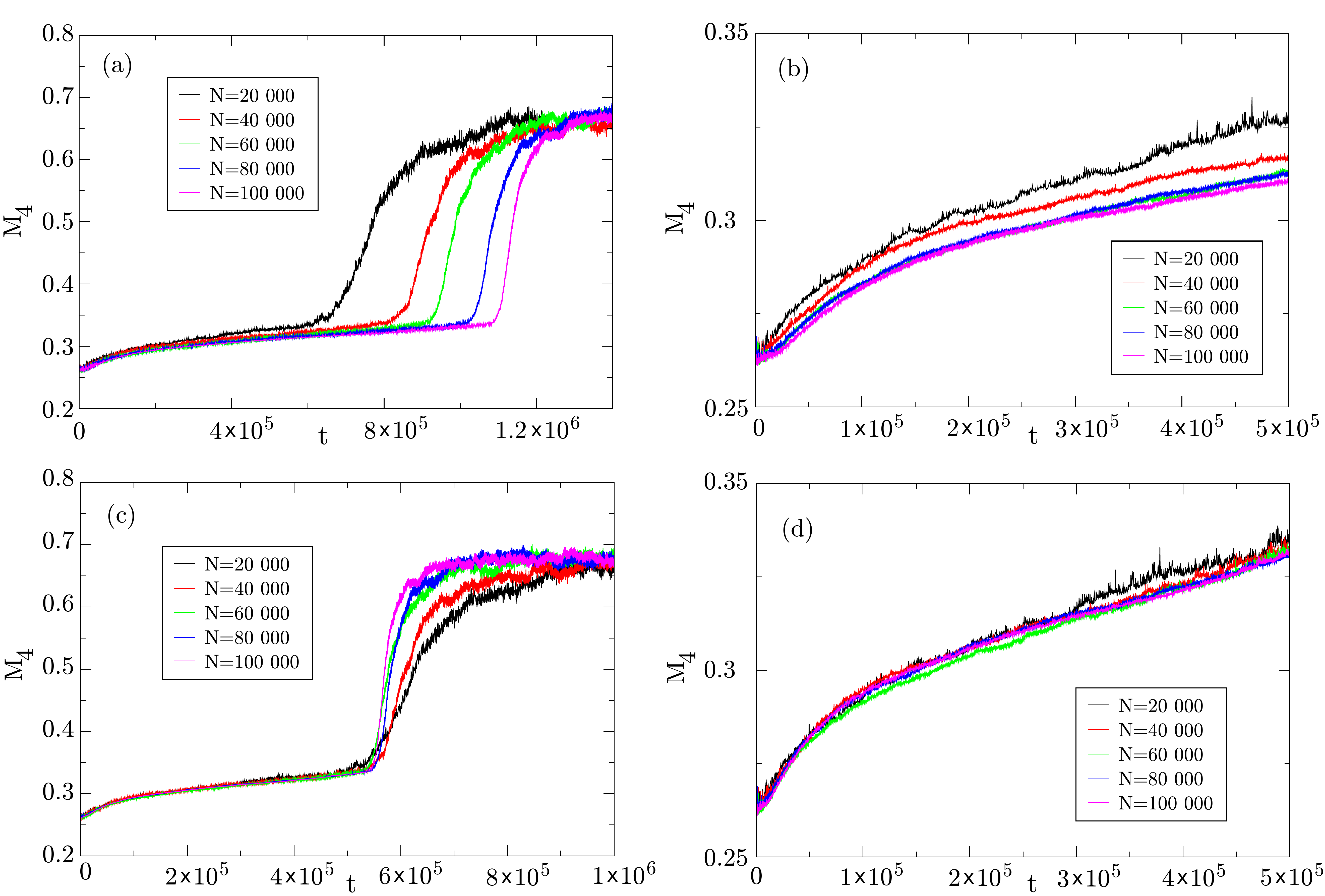}}}
\end{center}
\caption{(Color online) Moments $M_4$ as a function of time for the HMF model, with a homogeneous waterbag initial condition with energy per particle
$e=0.69$ and $N=20\,000$; $40\,000$; $60\,000$; $80\,000$; $100\,000$.
a) Time is rescaled as $t\rightarrow t/ (N\times 10^{-3})^{1.7}$. b) Time window corresponding to the duration of the QSS with
the same time scaling as (a). c) Time rescaled as $t\rightarrow t/ (N\times 10^{-3})^2$.
d) Time window corresponding to the duration of the QSS with the same time scaling as (a). The graphics shows clearly that for larger number of particles
the correct scaling is $N^2$.}
\label{fig2}
\end{figure}

\begin{figure}[ptb]
\begin{center}
\scalebox{0.3}{{\includegraphics{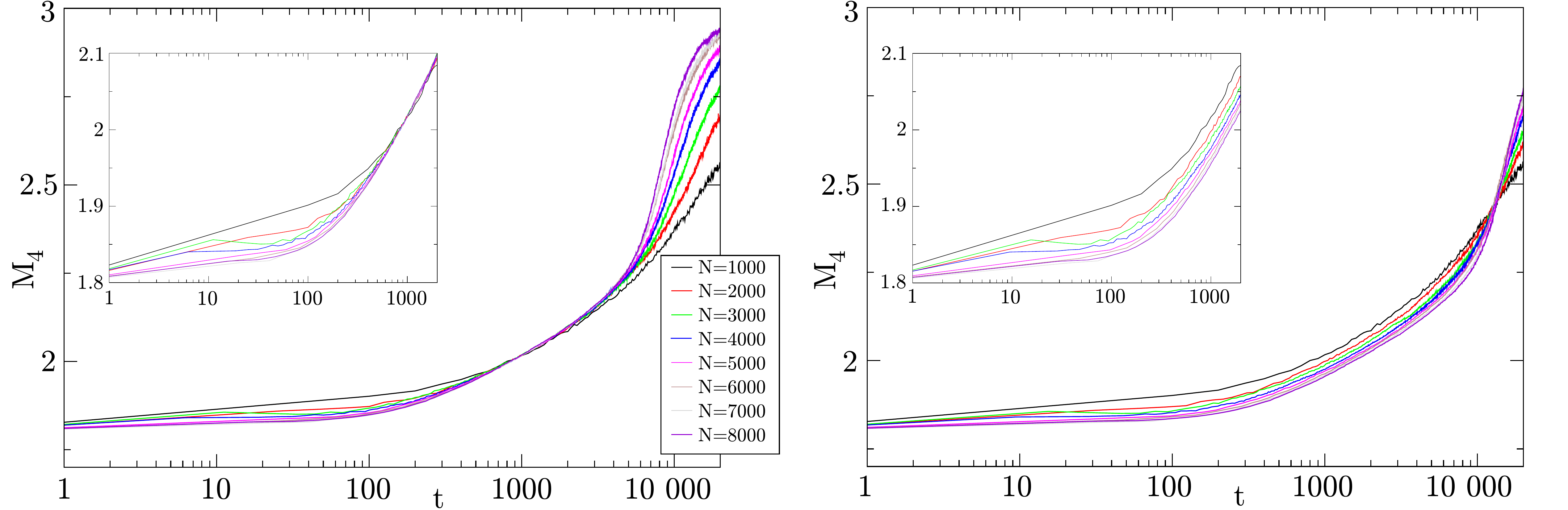}}}
\end{center}
\caption{(Color online) Moments $M_4$ as a function of time for the HMF model,
with a homogeneous waterbag initial condition with energy per particle
$e=0.69$ and small $N$ from $1\,000$ up to $8\,000$, averaged over $100$ realizations each.
Left Panel: the time was rescaled as $t\rightarrow t/ (N\times 10^{-3})^2$. Right Panel: time rescaled as
$t\rightarrow t/ (N\times 10^{-3})^{1.7}$. The insets are a zoom over the time interval where the system is
in a homogeneous state.}
\label{fig1}
\end{figure}

In order to develop our approach, we show in Fig.~\ref{fig2} the numerical results  for larger values of $N$, as the fluctuations turn out to be less important, and it can be seen that a $N^2$ scaling is yet more evident, while the $N^{1.7}$ scaling is clearly inappropriate. 
In Fig.~\ref{fig1} we show numerical results for the time evolution of $M_4$ for the same 
range of particle numbers $N$ as in Ref.~\cite{nv1} with two different time scalings: $N^{1.7}$ and $N^2$.
Although the influence of fluctuations are important, a somewhat better data collapse is obtained for the $N^2$ scaling.

\begin{figure}[ptb]
\begin{center}
\scalebox{0.7}{{\includegraphics{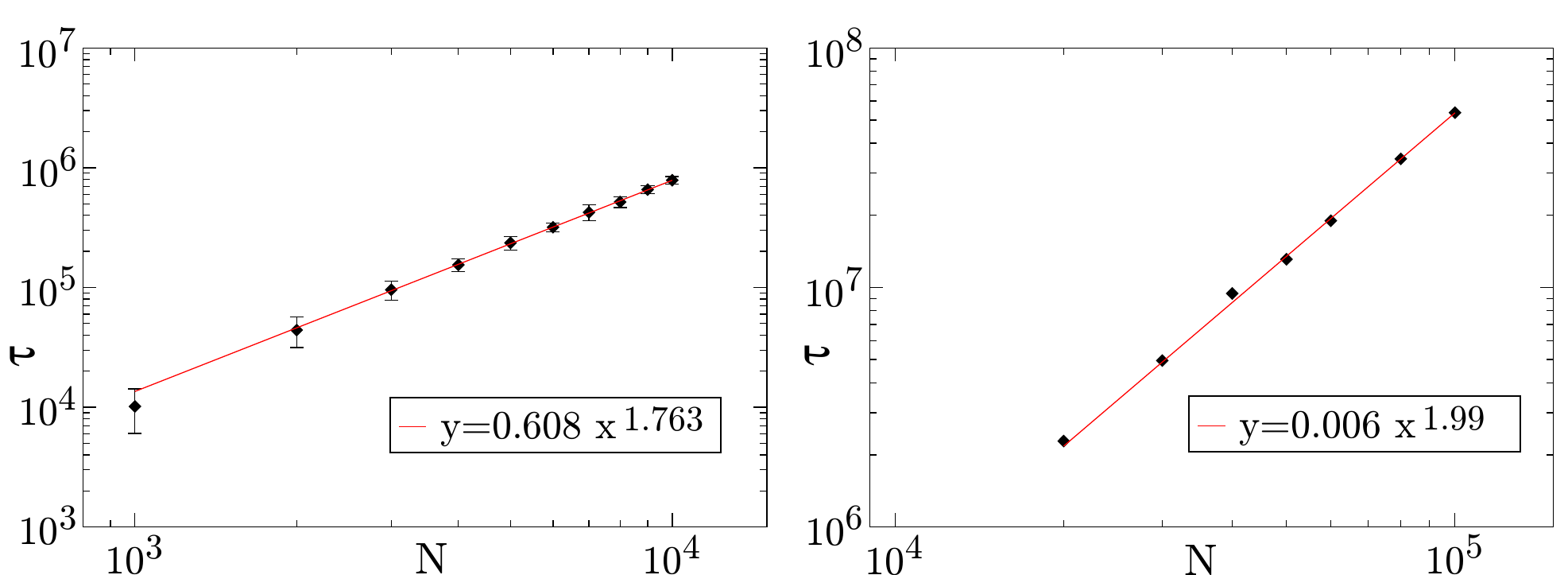}}}
\end{center}
\caption{(Color online) Left Panel: Life-times for the homogeneous waterbag state of the HMF model with energy per particle $e=0.69$ (solid circles)
with respective error bars. The best fit yields a power law scaling $N^{1.76}$.
Right Panel: same as the left panel but with greater number of particles and a best fit scaling of $N^{1.99}$. In this case the error bars are
at most the size of the circles.}
\label{fig4}
\end{figure}

Figure~\ref{fig4} shows the observed lifetimes of the QSS for the initial conditions
in Figs.~\ref{fig2} and~\ref{fig1}. The lifetimes are determined when the potential energy
varies more than a given percentage (5\%) of its QSS value.
In the left panel a scaling close to $N^{1.7}$ is obtained for $N<10\,000$ while
the $N^2$ scaling is obtained for $N>10\,000$ in the right hand panel. Note also that the error
bars for smaller $N$ are far from negligible such that a $N^2$ scaling is compatible with simulation data.
This apparent discrepancy is explained by noting that the life-time of the homogeneous QSS is defined by the exact moment in time that the states
looses its stability. Once the distribution function evolved closer to the stability threshold~\cite{artigo1}, fluctuations may trigger
an instability and drive the system out of the QSS regime. Therefore fluctuations play an important role for small $N$  (they are
of order $N^{-1/2}$), and thus generate a larger error bar for the life-times. In this case, the dynamics is governed by a kinetic equation with a $N^2$ scaling,
while the life-time can in average deviate from this scaling as a consequence of larger fluctuations for small $N$.

\begin{figure}[ptb]
\begin{center}
\scalebox{0.35}{{\includegraphics{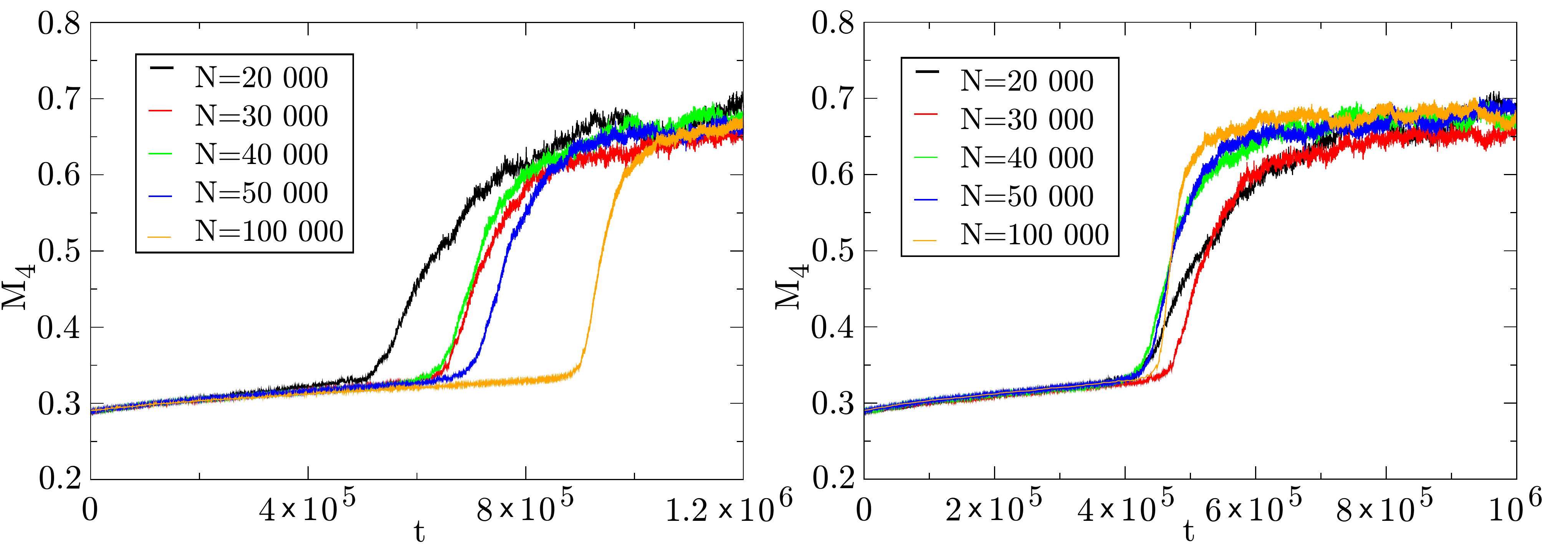}}}
\end{center}
\caption{(Color online) Moments $M_4$ as a function of time for the HMF model, with a homogeneous semi-elliptic initial condition with energy per particle
$e=0.69$ and $N=20\,000$; $30\,000$; $40\,000$; $50\,000$; $100\,000$.
Left Panel: the time was rescaled as $t\rightarrow t/ (N\times 10^{-3})^{1.7}$. Right Panel: time rescaled as
$t\rightarrow t/ (N\times 10^{-3})^2$.}
\label{fig3}
\end{figure}

\begin{figure}[ptb]
\begin{center}
\rotatebox{0}{\scalebox{0.3}{{\includegraphics{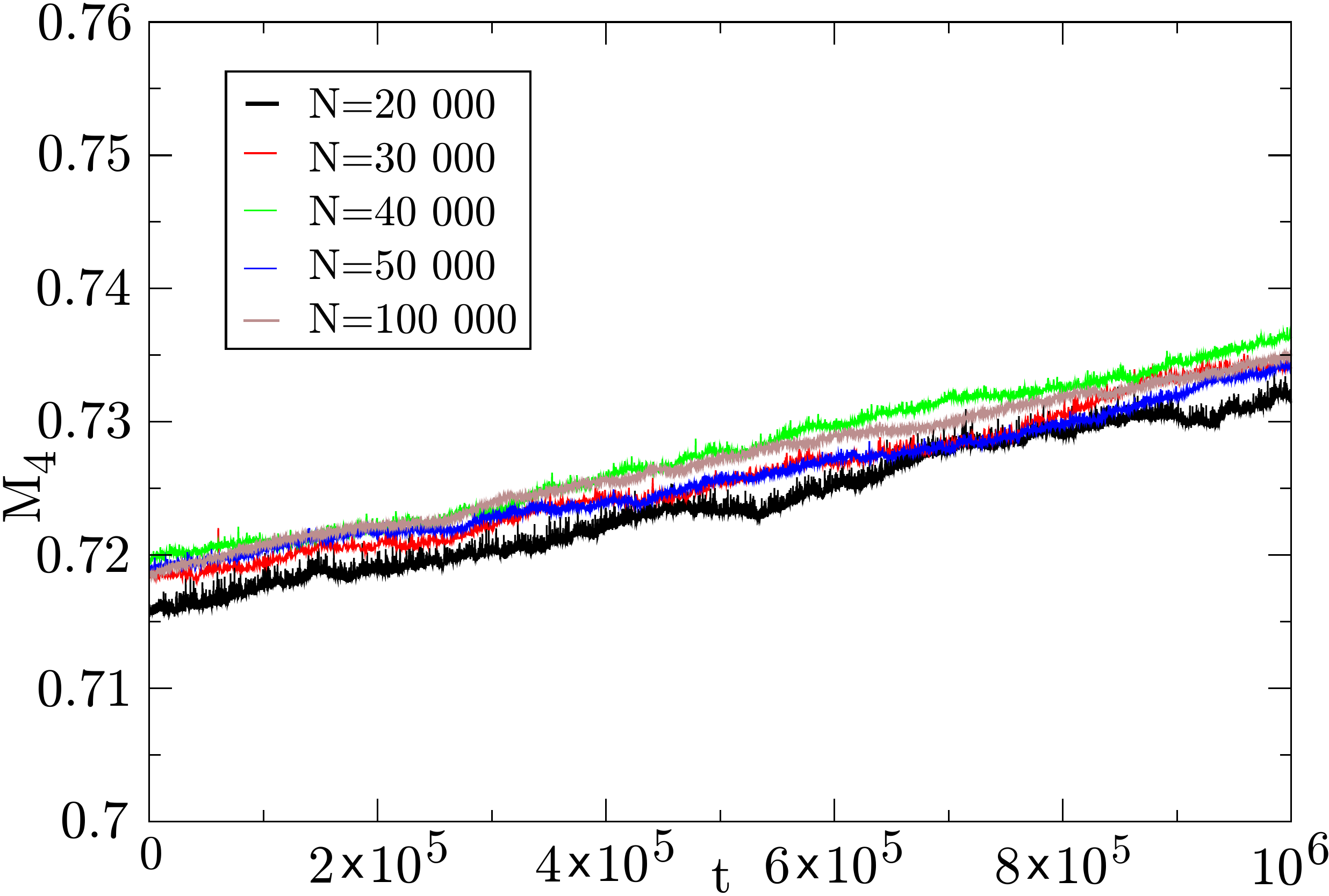}}}}
\end{center}
\caption{(Color online) Fourth moment $M_4$ of the velocity distribution as a function of time for the HMF model, with a homogeneous semi-elliptic initial condition with energy per particle
$e=0.8$ and $N=20\,000$; $30\,000$; $40\,000$; $50\,000$; $100\,000$. The time variable is rescaled as $t\rightarrow t/ (N/2\times 10^{-4})^{2}$ revealing the
data collapse.}
\label{fig5}
\end{figure}

To address the scaling $\exp(N)$ as presented in Ref.~\cite{nv2} for the HMF model we redo the same simulations under the same type of semi-elliptic
initial conditions. Figure~\ref{fig3} shows the time evolution
of $M_4$ for a time span greater than the lifetime of the QSS for $\epsilon=0.69$
and Fig.~\ref{fig5} shows results for $\epsilon=0.8$.
The latter case corresponds to the same energy used in Ref.~\cite{nv2} and is such that the system remains
always homogeneous up to the final thermodynamic equilibrium.
Again a very good data collapse is obtained for the $N^2$ scaling for both energies.

\begin{figure}[ptb]
\begin{center}
\rotatebox{0}{\scalebox{0.3}{{\includegraphics{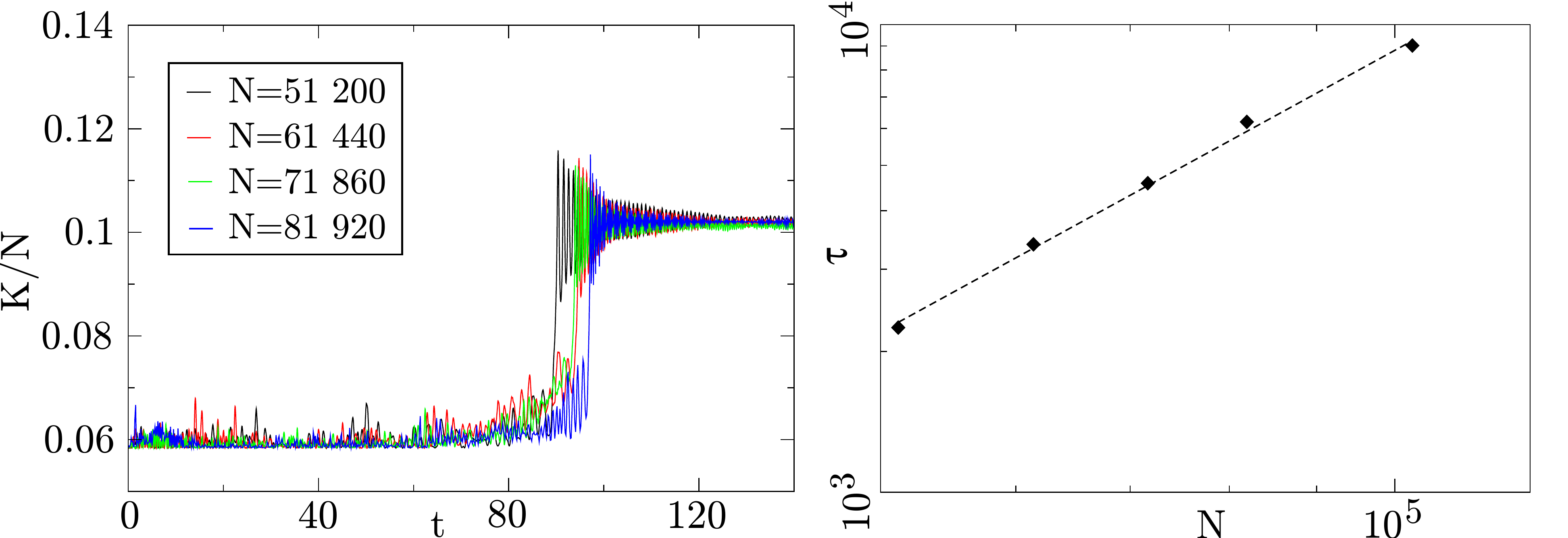}}}}
\end{center}
\caption{Left panel: (Color online) Kinetic energy per particle for the ring model for a homogeneous waterbag initial condition with $p_0=0.59$ ($f(p)=1/2p_0$ if $-p_0<p<p_0$) and
$\epsilon=0.1$, for a few different numbers of particles. The time is rescaled as $t\rightarrow t/ (N/10240)^2$.
Right panel: lifetimes for the QSS's in the left panel. The dashed
line corresponds to the $N^2$ scaling.}
\label{fig6}
\end{figure}

We also consider here the scaling with time of the lifetime of homogeneous QSS for the Ring model. The time dependence of the
kinetic energy for some values of $N$ and $\epsilon=0.1$ are shown in
Fig.~\ref{fig6}. We note that due to the QSS's being close to the stability threshold
fluctuations are more pronounced, but the $N^2$ scaling is nevertheless clearly observed.

All the numeric data presented  here points out to a kinetic equation obtained from the next order correction in $N^{-1}$.
To justify theoretically our findings we now briefly sketch
how to obtain a kinetic equation for the simpler HMF system, leaving a more detailed discussion
of a kinetic equation valid at the next order of approximation, either in a $1/N$ expansion or in the weak coupling limit characterized by parameter $\alpha$
to a separate publication~\cite{nosnew}. The HMF system (as the ring model) belongs to a class of systems with generic Hamiltonian
\begin{equation}
H(p,\theta)=\sum_{j=1}^N\frac{p_j^2}{2}+\frac{\alpha}{N}\sum_{j<k=1}^N V(\theta_j-\theta_k).
\label{genham}
\end{equation}
The pair potential interaction is rescaled by the Kac factor $1/N$ in such a way that
the energy is extensive (but non-additive) as for the Hamiltonians in Eqs.~(\ref{hmfham}) and~(\ref{ringham})~\cite{kac}.
The meaning of the dimensionless parameter $\alpha$ will be explained later in the paper.
The kinetic equation for a homogeneous state
can be obtained starting from the Liouville equation $\partial f_N/\partial t=\{H,f_N\}$ for the $N$ particle distribution function
in phase space $f_N(\theta_1,v_1,\ldots,\theta_N,v_N;t)$ normalized to unity and $\{H,f_N\}$ is the Poisson bracket of $f_N$ with $H$.
Defining the $s$-particle distribution function:
\begin{equation}
f_s\equiv f_s(\theta_1,v_1,\ldots,\theta_s,v_s;t)=\int \dd \theta_{s+1}\dd v_{s+1}\cdots \dd \theta_N\dd v_N\: f_N(\theta_1,v_1,\ldots,\theta_N,v_N;t).
\label{deffs}
\end{equation}
the BBGKY hierarchy is obtained as~\cite{balescu1,liboff}:
\begin{equation}
\frac{\partial}{\partial t} f_s=\sum_{j=1}^s\hat{K}_j f_s+\frac{1}{N}\sum_{j<k=1}^s\hat\Theta_{jk}f_s
+\frac{N-s}{N}\sum_{j=1}^s\int \dd \theta_{s+1}\dd v_{s+1}\hat\Theta_{j,s+1}f_{s+1},
\label{bbgky2}
\end{equation}
with
\begin{equation}
\hat{K}_j=-v_j\frac{\partial}{\partial \theta_j},\hspace{5mm}\hat\Theta_{jk}=-\frac{\partial}{\partial \theta_j}V(\theta_j-\theta_k)\partial_{jk},
\hspace{5mm}
\partial_{jk}\equiv\frac{\partial}{\partial v_j}-\frac{\partial}{\partial v_k}.
\label{defcalkth}
\end{equation}
From this point onward we replace $p_i$ by $v_i$ (all particles have unit mass). In order to obtain a kinetic equation for the one-particle distribution function
$f_1$ we must obtain $f_2$ as a functional of $f_1$. For that purpose we introduce the
irreducible cluster representation (correlation expansion) for the reduced
distribution functions $f_s$ up to $s=4$ as~\cite{balescu1,liboff}:
\begin{eqnarray}
 f_2(1,2) & = & f_1(1)f_1(2)+C_2(1,2),
\label{clusterrepa}
\\
 f_3(1,2,3) & = & f_1(1)f_1(2)f_1(3)+\sum_{P(i,j,k)} f_1(i)C_2(j,k)+C_3(1,2,3),
\label{clusterrepb}
\\
 f_4(1,2,3,4) & = & f_1(1)f_1(2)f_1(3)f_1(4)+\sum_{P(i,j,k,l)}\left[ f_1(i)C_3(j,k,l)+C_2(i,j)C_2(k,l)\right.
\nonumber\\
 & & \left.+f_1(i)f_1(j)C_2(k,l)\right]+C_4(1,2,3,4),
\label{clusterrepc}
\end{eqnarray}
where for simplicity the dependencies of each function on particle position and velocity is represented by the particle index,
e.~g.\ $C_2(1,2)\equiv C_2(\theta_1,\theta_2,v_1,v_2,t)$ and
$\sum_{P(i,j,k)}$ stands for the sum over all different permutations between particles $1$, $2$ and $3$, $f_1(1)\equiv f(v_1)$,
$f_2(1,2)\equiv f_2(\theta_1-\theta_2,v_1,v_2)$, $f_3(1,2,3)\equiv f_2(\theta_1-\theta_2,\theta_2-\theta_3,v_1,v_2,v_2)$ and similarly for the other terms.
Since the reduced distributions are taken
as fully symmetric by the permutation of any two particles, the same is valid for the pure correlation functions
$C_s$.

Two-particle correlation requires the interaction of
two particles, and therefore $C_2$ is of order $N^{-1}$. For three-particle correlations a two particle interaction between say particles 1 and 2, and then
between particles 2 and 3 are required, and $C_3$ is therefore of order $N^{-2}$, and so on.
The Vlasov and Landau equations are obtained at order $N^0$ and  $N^{-1}$, respectively. In order to explain our results
and since the collision term vanishes at order $N^{-1}$, we must resort to the next order term proportional to $1/N^2$ .
For a one-dimensional homogeneous system we obtain after replacing Eq.~(\ref{clusterrepa}) into Eq.~(\ref{bbgky2}):
\begin{equation}
\frac{\partial}{\partial t}f(v_1,t)=\int \dd \theta_2 \dd v_2\hat\Theta_{12}C_2(\theta_1,\theta_2,v_1,v_2,t).
\label{eqkin1}
\end{equation}
In the equation above we dropped out the index in $f_1$, and in the subsequent equations. We now expand $C_2$ and $C_3$ in the form
\begin{equation}
C_2=\frac{1}{N}C_2^{(1)}+\frac{1}{N^2}C_2^{(2)}+{\cal O}\left(\frac{1}{N^3}\right),\hspace{5mm}
C_3=\frac{1}{N^2}C_3^{(2)}+{\cal O}\left(\frac{1}{N^3}\right).
\label{c2series}
\end{equation}
Inserting the cluster expansion in Eqs.~(\ref{clusterrepa}--\ref{clusterrepc}) and the expansions given at Eq.~(\ref{c2series})
into the hierarchy in Eq.~(\ref{bbgky2}) we obtain for $s=2$ at orders $1/N$ and $1/N^2$, respectively:
\begin{eqnarray}
\lefteqn{\left(\frac{\partial}{\partial t}+v_1\frac{\partial}{\partial\theta_1}+v_2\frac{\partial}{\partial\theta_2}\right)
C_2^{(1)}(\theta_{12},v_1,v_2,t)=
V^\prime(\theta_{12})\:\partial_{12}f(v_1,t)f(v_2,t)}
\nonumber\\
 & & +\frac{\partial}{\partial v_1}f(v_1,t)\int \dd\theta_3\dd v_3\:V^\prime(\theta_{13})C_2^{(1)}(\theta_{23},v_2,v_3,t)
\nonumber\\
 & & +\frac{\partial}{\partial v_2}f(v_2,t)\int \dd\theta_3\dd v_3\:V^\prime(\theta_{23})C_2^{(1)}(\theta_{13},v_1,v_3,t),
\label{order-1}
\end{eqnarray}
\begin{eqnarray}
 & & \hspace{-10mm}\left(\frac{\partial}{\partial t} +v_1\frac{\partial}{\partial\theta_1}+v_2\frac{\partial}{\partial\theta_2}\right)
C_2^{(2)}(\theta_{12},v_1,v_2,t)=
\nonumber\\
 & & -f(v_1,t)\int\dd\theta_3\dd v_3\:V^\prime(\theta_{23})\frac{\partial}{\partial v_2}C_2^{(1)}(\theta_{23},v_2,v_3,t)
\nonumber\\
 & & -2\frac{\partial}{\partial v_1}f(v_1,t)\int\dd\theta_3\dd v_3\:V^\prime(\theta_{13})C_2^{(1)}(\theta_{23},v_2,v_3,t)
\nonumber\\
 & & +\frac{\partial}{\partial v_1}f(v_1,t)\int\dd\theta_3\dd v_3\:V^\prime(\theta_{13})C_2^{(2)}(\theta_{23},v_2,v_3,t)
\nonumber\\
 & & +\int\dd\theta_3\dd v_3\:V^\prime(\theta_{23})\frac{\partial}{\partial v_2}C_3^{(2)}(\theta_{12},\theta_{23},v_1,v_2,v_3,t)
+1\longleftrightarrow 2
\nonumber\\
 & & +V^\prime(\theta_{12})\:\partial_{12}C_2^{(1)}(\theta_{12},v_1,v_2,t),
\label{order-2}
\end{eqnarray}
with $1\longleftrightarrow 2$ standing for permutation of particles $1$ and $2$. For $s=3$ we obtain at the leading order:
\begin{eqnarray}
 & &\left(\frac{\partial}{\partial t} +v_1\frac{\partial}{\partial\theta_1}+v_2\frac{\partial}{\partial\theta_2}
+v_3\frac{\partial}{\partial\theta_3}\right)
C_3^{(2)}(\theta_{12},\theta_{23},v_1,v_2,v_3,t)=
\nonumber\\
 & & \sum_{P(i,j,k)}\left[\partial_{ij}V^\prime(\theta_{ij})C_2^{(1)}(\theta_{jk},v_j,v_k,t)
-f(v_i,t)\frac{\partial}{\partial v_j}f(v_j,t)\int\dd\theta_4\dd v_4 V^\prime(\theta_{j4})C_2^{(1)}(\theta_{k4},v_k,v_4,t))\right.
\nonumber\\
 & &
+\frac{\partial}{\partial v_i}f(v_i,t)\int\dd\theta_4\dd v_4\:V^\prime(\theta_{i4})C_3^{(2)}(\theta_{jk},\theta_{k4},v_j,v_k,v_4)
\nonumber\\
 & & \left.
+\frac{\partial}{\partial v_i}C_2^{(1)}(\theta_{ij},v_i,v_j,t)\int\dd\theta_4\dd v_4\:V^\prime(\theta_{j4})C_2^{(1)}(\theta_{j4},v_j,v_4,t)\right].
\label{s3}
\end{eqnarray}
In Eqs.~(\ref{order-1}--\ref{s3}) we used the notation $\theta_{ij}\equiv\theta_i-\theta_j$ as for a homogeneous state $f_s$ and $C_s$ depend on position only
through $\theta_{ij}$.

Equation~(\ref{order-1}) is and integro-differential equation for $C_2^{(1)}$. Noting that only the integral of $H_2\equiv\int\dd\:v_3 C_2^{(1)}$
is required in the kinetic integral, Lenard was able to solve the corresponding equation for $H_2$, but not for $C_2^{(1)}$ explicitly~\cite{lenard},
which is required to solve the remaining equations~(\ref{order-2}) and~(\ref{s3}).
With some additional considerations on the strength of the inter-particle force in a homogeneous state
a closed form expression for the two and three-particle correlation functions can be obtained, as we proceed to show. 

\begin{figure}[ptb]
\begin{center}
\scalebox{0.25}{{\includegraphics{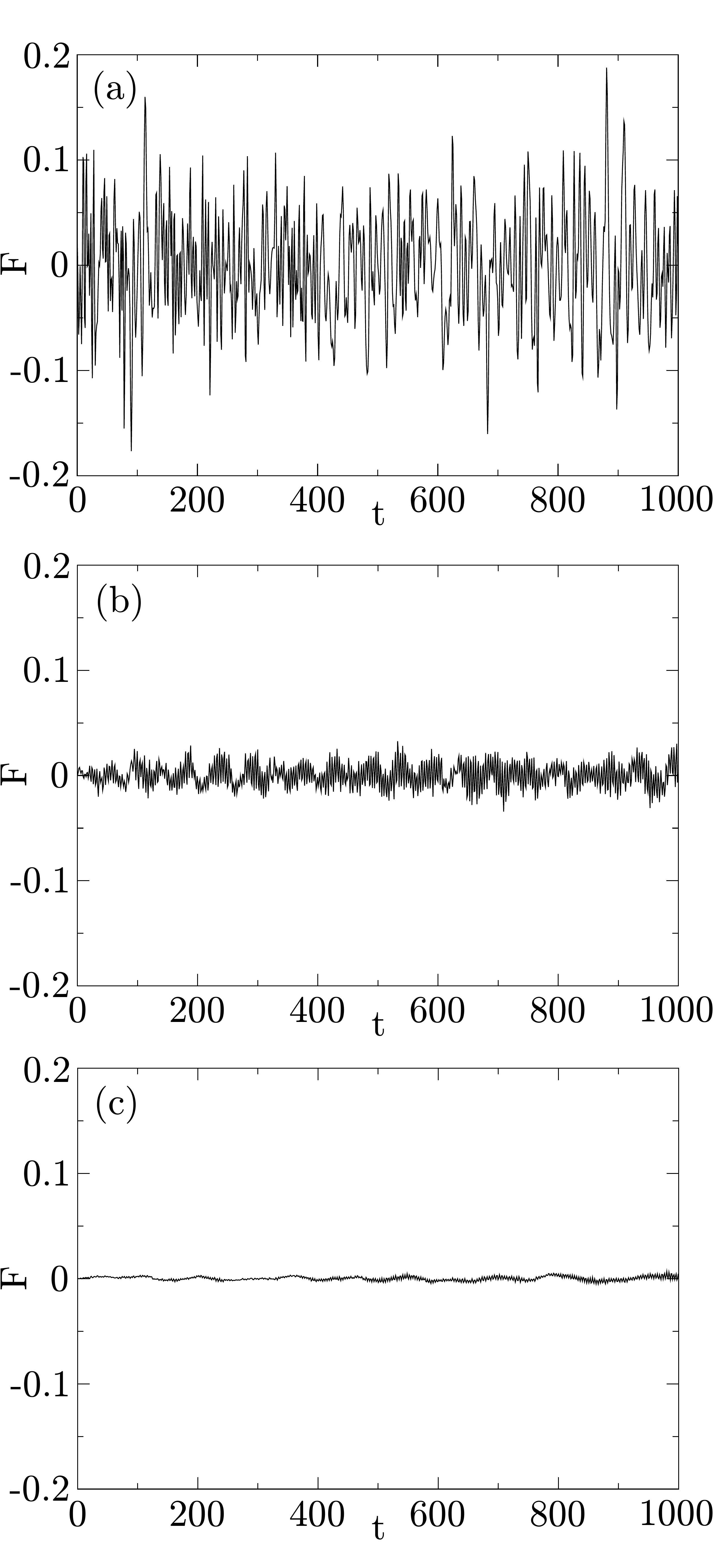}}}
\end{center}
\caption{Force on a single particle for a)~$N=1000$, b)~$N=10\,000$ and c)~$N=10\,000\,000$ with energy per particle $e=0.69$.}
\label{fig7}
\end{figure}

We observe that in a homogeneous state of the HMF model the force on a given particle is very small. Thus
it is reasonable on a phenomenological ground to use some sort of weak coupling approximation to simplify the system of Eqs.~(\ref{order-1}--\ref{s3}).
Indeed the mean-field force on each particle is given by $F=-V^\prime\propto (1/N)\times N\times\alpha=\alpha$, i.~e.\ the Kac factor times the particle number
(from the sum over the remaining particles)
with $\alpha$ characterizing the effective interaction strength in the homogeneous state which is proportional to
the small magnitude of the fluctuations of the space distribution function around homogeneity.
Such fluctuations, as illustrated in Fig.~\ref{fig7}, diminishes with increasing $N$.
In fact, we are considering that the approximation presented is a generalization of the Landau Equation for one-dimensional systems in a homogeneous state.
In this way we have the following orders of magnitude: $C_s\propto\alpha^{s-1}$ and $V^\prime\propto\alpha$.
In Eq.~(\ref{order-1}) the last two terms on the right-hand side can then be neglected, i.~e.\ we keep terms up to order $\alpha$ and neglect terms of order $\alpha^2$.
The third term in the summation in the right-hand side of Eq.~(\ref{s3}) is of order $\alpha^3$ and is also negligible.

Since the physical space of the HMF model is periodic,
it is convenient to introduce the Fourier series (discarding for convenience the constant term in $V$):
\begin{eqnarray}
 & & V(\theta_{12})=\sum_n \tilde V(n){\rm e}^{in\theta_{12}},\hspace{5mm}C_2(\theta_{12},v_1,v_2;t)=\sum_n \tilde{C}_2(k,v_1,v_2;t){\rm e}^{in\theta_{12}},
\nonumber\\
 & & C_3(\theta_{12},\theta_{23},v_1,v_2;t)=\sum_{n,m} \tilde{C}_3(n,m,v_1,v_2;t){\rm e}^{in\theta_{12}}{\rm e}^{im\theta_{23}}
\label{fourseries}
\end{eqnarray}
with $n,m$ integers ranging from $-\infty$ to $\infty$ and
\begin{eqnarray}
 & & \tilde V(n)=\frac{1}{2\pi}\int\dd\theta_{12} V(\theta_{12}){\rm e}^{-in\theta_{12}}=-\frac{1}{2}\left[\delta_{n,1}+\delta_{n,-1}\right],
\nonumber\\
 & & \tilde C_2(n,v_1,v_2;t)=\frac{1}{2\pi}\int \dd \theta_{12} C_2(\theta_{12},v_1,v_2;t){\rm e}^{-in\theta_{12}},
\nonumber\\
 & & \tilde C_3(n,m,v_1,v_2,v_3;t)=\frac{1}{(2\pi)^2}\int\dd\theta_{12}\dd\theta_{23}C_2(\theta_{12},\theta_{23},v_1,v_2,v_3;t)
{\rm e}^{-in\theta_{12}}{\rm e}^{-in\theta_{23}},
\label{vcfourier}
\end{eqnarray}
where $\delta_{n,m}$ is the Kronecker delta.
Note that $C_4$ is neglected at this approximation as it is of order $1/N^3$.  Equation~(\ref{eqkin1}) is rewritten as
\begin{eqnarray}
\frac{\partial}{\partial t} f(v_1;t) & = & i\int \dd v_2\sum_n n\tilde V(n)\:\partial_{12}\tilde{C}_2(n,v_1,v_2;t)
\nonumber\\
 & = & \frac{i}{2}\frac{\partial}{\partial v_1}\int \dd v_2\left[\tilde C_2(-1,v_1,v_2,t)-C_2(1,v_1,v_2,t) \right]
\nonumber\\
 & = & \frac{\partial}{\partial v_1}\int \dd v_2\:\im\left[C_2(1,v_1,v_2,t)\right],
\label{eqkin1b}
\end{eqnarray}
where we used the property $C_2(-1,v_1,v_2,t)=C_2(1,v_1,v_2,t)^*$ as it is the Fourier coefficient of a real function.
Thus only the imaginary part of the Fourier transform $\tilde C_2$ contributes to the kinetic equation.

Using the Fourier series in Eq.~(\ref{fourseries})
and~(\ref{vcfourier}) as well as the Fourier coefficients for the HMF potential yields:
\begin{eqnarray}
\left(\frac{\partial}{\partial t}+iv_{12}m\right)\tilde C_2^{(1)}(m,v_1,v_2,t)=-\frac{i}{2}\left(\delta_{m,1}-\delta_{m,-1}\right)
\partial_{12}f(v_1,t)f(v_2,t),
\label{eqc21k}
\end{eqnarray}
with $v_{jk}\equiv v_j-v_k$ and $\tilde C_n^{(s)}$ the Fourier coefficient of $C_n^{(s)}$.
Before solving Eq.~(\ref{eqc21k}) we must determine the time evolution of the one-particle distribution function,
which is given by the ballistic approximation (free motion with constant velocity), valid up to order $1/N^2$  in the present case $f(v,t)=f(v)+{\cal O}(N^{-2})$.
The solution of Eq.~(\ref{eqc21k}) is now easily obtained as:
\begin{eqnarray}
\tilde C_2^{(1)}(m,v_1,v_2,t) & = & \tilde C_2^{(1)}(m,v_1,v_2,0)\:{\rm e}^{-iv_{12}mt}
\nonumber\\
 & & -\frac{i}{2}\left(\delta_{m,1}-\delta_{m,-1}\right)\int_0^t\dd\tau\:{\rm e}^{iv_{12}m\tau}
\partial_{12}f(v_1,\tau)f(v_2,\tau).
\label{soleqc21k}
\end{eqnarray}
The first term on the right-hand side is a transient that describes the memory of the initial correlation and rapidly becomes negligible
(see~\cite{balescu1} for a detailed discussion of this point). On the other hand, the integral in $\tau$ tends to the
Cauchy integral for large $t$~\cite{balescu1}:
\begin{equation}
\int_0^\infty \dd t\: {\rm e}^{iv_{12}mt}=\pi\delta(v_{12}m)+i{\cal P}\left(\frac{1}{v_{12}m}\right),
\label{cauchyint}
\end{equation}
with ${\cal P}(1/a)$ the principal part of $1/a$ and $\delta(a)$ the Dirac delta function.
This last step is a Markovianization procedure of the solution for the correlation function.
Therefore, the solution of Eq.~(\ref{order-1}) takes the form:
\begin{equation}
\tilde C_2^{(1)}(m,v_1,v_2,t)=\frac{1}{2}{\cal P}\left(\frac{1}{v_{12}m}\right)\left(\delta_{m,1}-\delta_{m,-1}\right)
\partial_{12}f(v_1,t)f(v_2,t).
\label{soleqc21kb}
\end{equation}
In Eq.~(\ref{order-2}) all terms must be kept. Indeed if we keep only terms of order $\alpha^2$ this would imply discarding all terms containing
$\tilde C_3^{(2)}$. It can be shown that the next order correction $\tilde C_2^{(2)}$ would be real (see for instance the appendix in Ref.~\cite{lenard}), not contributing
to the kinetic equation. Consequently the next order non-vanishing correction to the kinetic equation comes from the contribution of three-particle correlations.
The later can be determined by solving Eq.~(\ref{s3}) neglecting terms of order $\alpha^3$, i.~e. the last two terms between brackets on the right-hand side.
Both remaining equations Eqs.~(\ref{order-2}) and~(\ref{s3}) are then written in Fourier space, and the latter solved for $\tilde C_3^{(2)}$. This solution
is then replaced in Eq.~(\ref{order-2}) and that is solved for $\tilde C_2^{(2)}$, which together with Eq.~(\ref{eqkin1b}) yield
the desired kinetic equation. All these calculations are straightforward but quite long and tedious, but easily handled
using a computer algebra system~\cite{maple} with specific routines developed by the authors for this purpose
and is given in Supplemental Material at [URL to be inserted by publisher].
Here we only show the final result for $\im\,\tilde C_2^{(2)}$ obtained as delineated above:
\begin{eqnarray}
\lefteqn{\im\,\tilde C_2(1,v1,v2,t)=\frac{1}{N^2}\im\,\tilde C_2^{(2)}(1,v1,v2,t)= \frac{\pi^2}{4}{\cal P}\left(\frac{1}{v_1-v_2}\right)}
\nonumber\\
 & & \hspace{-5mm}\times\left.
\left\{
{\cal P}^\prime\left(\frac{1}{v_1-v_2}\right)\hat{\cal K}_I+{\cal P}^{\prime\prime}\left(\frac{1}{v_1-v_2}\right)\hat{\cal K}_{II}
\right\}f(v_1)f(v_2)f(v_3)\right|_{v_3=2v_2-v_1}+\:\: 1\longleftrightarrow 2,
\label{c22kfinal}
\end{eqnarray}
where $1\longleftrightarrow 2$ stands for terms obtained from a permutation of particles $1$ and $2$, and
\begin{eqnarray}
 & & \hat{\cal K}_I\equiv 3\frac{\partial}{\partial v_2}\frac{\partial}{\partial v_3}-2\frac{\partial^2}{\partial v_3^2}
-2\frac{\partial}{\partial v_1}\frac{\partial}{\partial v_3}-\frac{\partial}{\partial v_1}\frac{\partial}{\partial v_2}
+2\frac{\partial^2}{\partial v_2^2},
\nonumber\\
 & & \hat{\cal K}_{II}\equiv -2\frac{\partial}{\partial v_2}+\frac{\partial}{\partial v_1}+\frac{\partial}{\partial v_3}.
\label{oprskdef}
\end{eqnarray}
Equations~(\ref{eqkin1b}) and~(\ref{c22kfinal}) give the final form of the kinetic equation for the HMF model in a homogeneous state.
The collisional integral is non-vanishing and proportional to $1/N^2$ explaining, from a theoretical viewpoint, the scale $N^2$ for
the dynamics of one-dimensional homogeneous systems, as shown in all the simulations reported in this paper. Our results are
at variance with previous results reported in the literature, and its origin may be understood as an effect of the size of $N$ on those calculations.
In Ref.~\cite{ettoumi} Ettoumi and Firpo developed a stochastic theory based on the first passage time approach and taking into account
the role of fluctuations. Their approach nevertheless requires the numeric determination
of a diffusion coefficient and is therefore limited to the number of particles considered (up to $N=20\,000$).
These authors were able to obtain the exponent $1.7$, and it
would be interesting to extend their work to larger number of particles as they have been considered here, where a $N^2$ scaling is expected.
Our results strongly suggest that although the dynamics, as represented here by the time evolution of the velocity moments, scales
as $N^2$, even for small $N$, but when the effect of fluctuations is taken into account in the destabilization of the QSS, a different
exponent can be obtained. Further research in course should clarify these points.
As a concluding remark, in the present paper we show that the use of the magnetization as relevant dynamical variable only depends on the space fluctuations around homogeneity
(the magnetization is always zero up to fluctuations). On the other hand,
the dynamics in the velocity distribution, which changes over time according to the kinetic equation, is not probed.
The statistical moments of the velocity distribution are therefore more suitable variables to describe the dependence of the dynamics
on the particle number $N$.

The authors acknowledge partial financial support by CAPES and CNPq (Brazilian agencies). TMRF would like to thank Bruno Marcos for fruitful discussions.


\begin{thebibliography}{}
\bibitem{lb1} This term was originally coined by LyndenBell who first developed a statistical theory for predicting its outcome.
\end{thebibliography}
\end{document}